\documentclass[letter]{aa}  

\usepackage{graphicx}
\usepackage{txfonts}
\usepackage{natbib,twoopt}
\usepackage[breaklinks=true]{hyperref} 
\bibpunct{(}{)}{;}{a}{}{,}             
\makeatletter
  \newcommandtwoopt{\citeads}[3][][]{\href{http://adsabs.harvard.edu/abs/#3}%
    {\def\hyper@linkstart##1##2{}%
     \let\hyper@linkend\@empty\citealp[#1][#2]{#3}}}
  \newcommandtwoopt{\citepads}[3][][]{\href{http://adsabs.harvard.edu/abs/#3}%
    {\def\hyper@linkstart##1##2{}%
     \let\hyper@linkend\@empty\citep[#1][#2]{#3}}}
  \newcommandtwoopt{\citetads}[3][][]{\href{http://adsabs.harvard.edu/abs/#3}%
    {\def\hyper@linkstart##1##2{}%
     \let\hyper@linkend\@empty\citet[#1][#2]{#3}}}
  \newcommandtwoopt{\citeyearads}[3][][]%
    {\href{http://adsabs.harvard.edu/abs/#3}
    {\def\hyper@linkstart##1##2{}%
     \let\hyper@linkend\@empty\citeyear[#1][#2]{#3}}}
\makeatother
\begin{document}

   \title{Survey of Orion Disks with ALMA (SODA)}

   \subtitle{II. UV-driven disk mass loss in L1641 and L1647}

   \author{S.E. van Terwisga
          \inst{1}
          \and
          A. Hacar\inst{2}
          }

   \institute{Max-Planck-Institut f\"{u}r Astronomie, K\"{o}nigstuhl 17, Heidelberg
              \email{terwisga@mpia.de}
              \and 
              Institute for Astronomy (IfA), University of Vienna,
              T\"urkenschanzstrasse 17, A-1180 Vienna\\
              }

   \date{\today}

\abstract{External FUV irradiation of protoplanetary disks has an important impact on their evolution and ability to form planets. However, nearby ($<300$\,pc) star-forming regions lack sufficiently massive young stars, while the Trapezium Cluster and NGC 2024 have complicated star-formation histories and their O-type stars' intense radiation fields ($>10^4\,G_0$) destroy disks too quickly to study this process in detail.}{We study disk mass loss driven by intermediate ($10 - 1000\,G_0$) FUV radiation fields in L1641 and L1647, where it is driven by more common A0 and B-type stars.}{Using the large (N=873) sample size offered by the Survey of Orion Disks with ALMA (SODA), we search for trends in the median disk dust mass with FUV field strength across the region as a whole and in two separate regions containing a large number of irradiated disks.}{For radiation fields between $1 - 100\,G_0$, the median disk mass in the most irradiated disks drops by a factor $\sim 2$ over the lifetime of the region, while the 95th percentile of disk masses drops by a factor 4 over this range. This effect is present in multiple populations of stars, and localized in space, to within $2$\,pc of ionizing stars. We fit an empirical irradiation - disk mass relation for the first time: $M_{\rm{dust,median}} = -1.3^{+0.14}_{-0.13} \log_{10}(F_{\rm{FUV}} / G_0) + 5.2^{+0.18}_{-0.19}$.}{This work demonstrates that even intermediate FUV radiation fields have a significant impact on the evolution of protoplanetary disks.}
 
   \keywords{protoplanetary disks}

   \maketitle
%

\section{Introduction}
The planetary systems known to date represent the outcome of planet formation in diverse star-forming environments. On the other hand, our understanding of the protoplanetary disks in which young planetary systems are forming is predominantly driven by observations of low-mass star-forming regions (SFRs) within 300\,pc, with typically only about $\sim 100$ young ($\sim1 - 10$\,Myr) disk-bearing stars. As such, these regions do not represent the more massive, clustered environments in which most stars form~\citep{lada03,bressert10}. The median star in the galaxy is formed in a region with a higher surface density of young stars, and -- as a result of the stellar mass distribution -- will be within a few pc of an O- or B-type young star at some point in its early life~\citep{winter22}. These young, massive stars affect the survival of disks in nearby stars through photo-evaporation, as shown by the presence of proplyds~\citep{odell93} and by theoretical work~\citep[e.g.,][]{scally01, FRIED, concharamirez19,emsenhuber23}. In the immediate proximity of O-type stars, EUV photons drive mass loss (for instance, within 0.3\,pc of $\theta^1$ Ori C (O6; \citealt{stoerzer99}). FUV takes over at larger distances and is the dominant cause of disk mass loss for lower-mass stars. Understanding the influence of more representative environments on the evolution of disks is therefore essential for connecting disk evolution to planetary system formation.  

As the nearest star-forming region with sufficiently high surface densities and several O-type stars, the Orion A and B molecular clouds are a crucial touchstone for models of disk destruction through external photoevaporation. Several Atacama Large Millimeter/submillimeter Array (ALMA) surveys of the cold millimeter-sized dust emission from disks in the Orion Nebula Cluster (ONC)~\citep{mann14,eisner18,vanterwisga19}, the $\sigma$ Ori region~\citep{ansdell17} and NGC 2024~\citep{vanterwisga20} have looked for disk mass loss driven by external FUV irradiation. In the ONC, disk masses appear to decrease with proximity to the O6 star $\theta^1$ Ori C, at least within 1\,pc; a decrease of detection rates and maximum disk masses is seen within 1\,pc of $\sigma$ Ori (O9;\citealt{maizapellaniz21}), while any trend in NGC 2024 is ambiguous.

\begin{figure*}[t]
    \centering
    \includegraphics[width=\textwidth]{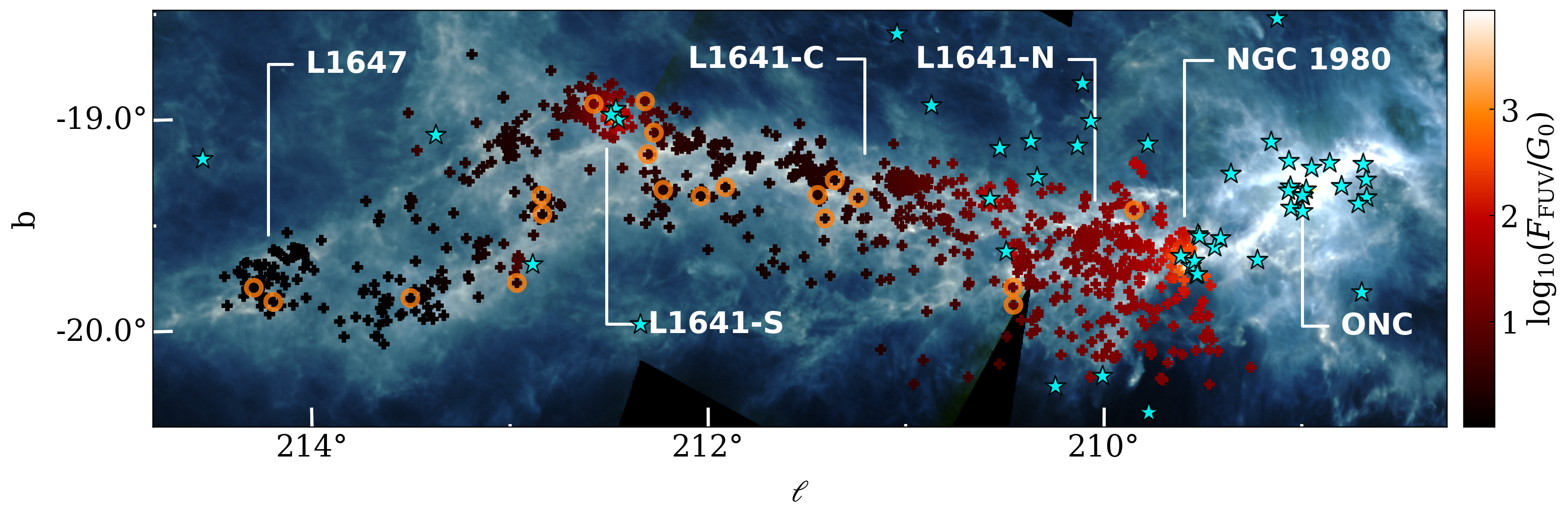}
    \caption{External FUV flux estimated for the SODA sample. Crosses indicate the positions of Class II disks in the SODA catalog: color indicates FUV irradiation. Disks with masses $> 100\,M_{\oplus}$ based on their millimeter flux are additionally marked with orange circles. Cyan stars mark possible ionizing stars in their vicinity. Background: \textit{Herschel} SPIRE observations at 250, 300, and 500\,$\mu$m.}
    \label{fig:G0map}
\end{figure*}

Interpreting these results remains challenging. All these surveys contain relatively small numbers of stars ($N \sim 100$). In the ONC, the observed disk mass effect may be due to the local history of star formation~\citep{winter19}. The interpretation of disk mass loss due to external photoevaporation was further cast into doubt by~\citet{parker21}, whose simulations suggest that low-number statistics in the inner regions of these regions might drive the observed trend, and who do not predict a distance-dependence of the disk mass due to the relative motions of stars in young clusters. More sophisticated models of young clusters by~\citet{concharamirez21, concharamirez22} likewise suggest the ONC disks are the result of ongoing star formation and dispersal through the cluster. From an observational point of view, in environments like the ONC, the typically assumed relations between disk luminosity and mass -- which imply optically-thin dust emission with disk radii of $\sim 60$\,AU and low midplane temperatures of 20\,K -- almost certainly do not apply~\citep{haworth21}.

In this Letter, we study the impact of intermediate FUV radiation fields ($10 - 1000\,G_{0}$, where $G_0$ is the Habing radiation field\footnote{$G_0 = 1.6 \times 10^{-3}$\,erg\,cm$^{-2}$\,s$^{-1}$, \citep{habing68}}) on the luminosity of a large sample of protoplanetary disks at millimeter wavelengths in L1641 and L1647. Unlike the regions studied so far, in this part of Orion A surface densities are lower ($\sim 10 - 100$\,pc$^{-2}$), ionization is primarily driven by A0 and B-type stars and is thus less efficient, meaning disks are not completely destroyed by 1\,Myr. The age structure is less ambiguous thanks to existing Gaia observations~\citep{kounkel18,zari19}. To compensate for the weaker expected effect, we rely on the large number ($N=873$) of disk-bearing stars characterized with ALMA in the Survey of Orion Disks with ALMA (SODA) by~\citet{SODA}. Thus, we can provide an unbiased view of the importance of external UV-driven evolution of disks in such - very common - environments for the first time.

\section{Methods}
The input data for this Letter consist of the SODA catalog of disk masses along the southern part of the Orion A cloud, from L1641 to L1647, covering more than 50\,pc in length. This survey, the largest of its kind so far, observed the mm-continuum flux of 873 protoplanetary disk candidates, identified by~\citet{megeath12} based on \textit{Spitzer} photometry. It is unresolved (with $1.4''$ resolution) but deep, with an rms of 0.08\,mJy\,beam$^{-1}$ or 1.5\,$M_{\oplus}$ at the $4\sigma$ level. For an in-depth discussion of the data processing and the catalog itself, see~\citet{SODA}.

We combine this survey with a catalog of stars sufficiently hot and luminous to drive external photoevaporation, with spectral types A0 or earlier, and sufficiently close (with distances $300-475$\,pc) to be associated with the Orion molecular cloud~\citep{grossschedl18,grossschedl21}. The distance uncertainties for these stars tend to be large as \textit{Gaia} data for stars of these magnitudes tends to be of low quality. We therefore limit ourselves to projected separations.

To estimate the candidate ionizing stars' FUV luminosity, we use BHAC-15 isochrones~\citep{baraffe15} and integrate the Castelli and Kurucz model spectra~\citep{CK04} of stars of similar effective temperatures between $911.6 - 2066$\,\AA. We do not take into account interstellar extinction, given the uncertain geometry, and stars beyond 10\,pc from a given disk are not included in the calculation. We assume a minimum interstellar FUV flux of 1\,$G_0$. The total irradiation per source in the SODA sample is calculated by adding the contribution of all O, B, and A0 stars within 10\,pc of each disk after diluting their emission for the projected separation, at the distance of each disk-bearing star. For these distances, we make the same assumptions as in~\citet{SODA}, and thus follow the large-scale cloud structure from~\citep{grossschedl18}.

For each Class II source in SODA, Figure~\ref{fig:G0map} shows the calculated incident external FUV flux. These range between $10^3\,G_0$ in the proximity of the NGC 1980 and L1641-S clusters, but falls to the standard $1\,G_0$ in the southernmost parts of L1647. All intermediate FUV regimes are well-sampled, but it is clear that massive stars are not distributed equally throughout the cloud. Interestingly, in L1641-S, L1641-N, and NGC 1980, which contain the most irradiated disk samples, the most massive disks (indicated with orange circles) appear to be less common even by eye. In the following sections, we investigate this observation systematically.

\section{Disk mass as a function of separation and irradiation}

\begin{figure}[h]
    \centering
    \includegraphics[width=\linewidth]{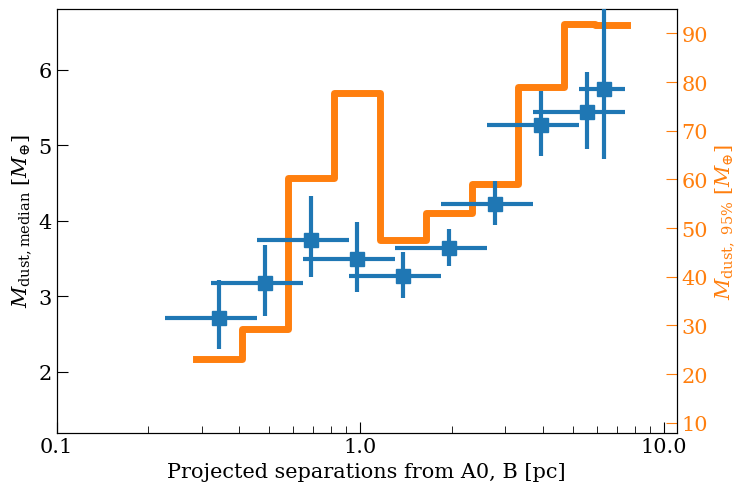}
    \caption{Disk masses binned by separation to potential ionizing stars. Blue points mark the median disk mass in each bin, with error bars indicating the statistical uncertainty after fitting a log-normal distribution. The orange line shows the empirical 95th percentile of disk masses in each bin.}
    \label{fig:M_vs_dist}
\end{figure}

First, we investigate how the median disk dust mass depends on the UV radiation field in the simplest sense, by considering only the separation to ionizing stars, like in previous work~\citep[e.g.][]{mann14, eisner18}. Following~\citet{SODA}, we exploit the result that protoplanetary disk flux (and mass) distributions are well-described by a log-normal distribution, with very similar widths between SFRs, but different median values. We sort and bin disk-bearing stars by separation from the nearest ionizing stars and calculate the median mass in each bin if it has $N > 50$ stars. The result is shown in Figure~\ref{fig:M_vs_dist}.

Clearly, disk masses are significantly lower within 0.5\,pc separation of an ionizing star, and this result holds both for the median disk mass (which differs by a factor 3 across the figure) and for the 95th-percentile mass (which shows an even stronger decrease). An increase in the 95th percentile disk mass in the 0.6 - 1\,pc range is surprising, and discussed in Sect.~\ref{sec:zoom}. Any trend beyond 0.5\,pc, however, is less (statistically) clear, with the most isolated disks apparently having higher masses, corresponding to L1647.

\begin{figure}[h]
    \centering
    \includegraphics[width=\linewidth]{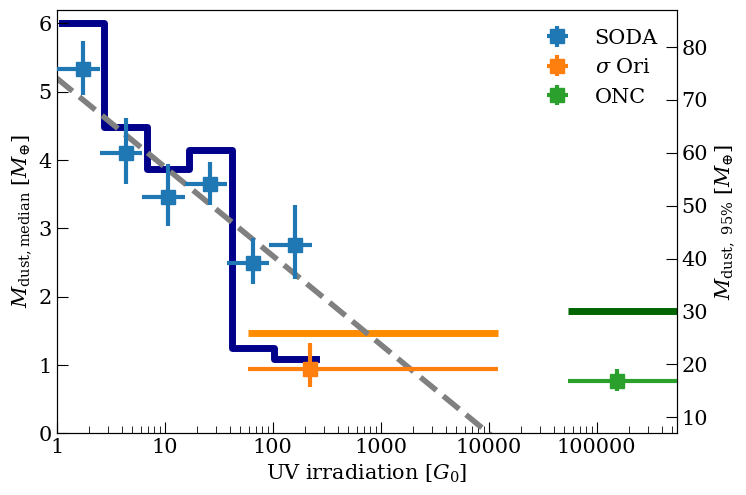}
    \caption{Disk masses binned by incident FUV flux (in units of $G_0$). Blue points show median dust masses in each bin, with uncertainties; the empirical 95th percentile disk mass is shown as a bold, dark blue line. The grey dashed line represents an empirical fit to the relation between the median dust mass and the FUV radiation field for the SODA sample. The points in orange and green represent, for $\sigma$ Ori and the ONC respectively, the median and $95\%$ disk mass (as lines), for the central $90\%$ of the FUV distribution in these regions.}
    \label{fig:M_vs_G0}
\end{figure}

As Figure~\ref{fig:G0map} shows, stars of different spectral types are driving external photoevaporation in L1641. Therefore, in Figure~\ref{fig:M_vs_G0} the median disk mass and the empirical 95th percentile disk mass are plotted relative to the (upper limit to the) external $F_{\rm{FUV}}$ in units of $G_0$. Once again, the median disk mass decreases significantly with external irradiation. By sampling from the distribution of median disk masses in each $F_{\rm{FUV}}$ bin and their uncertainties, we find a Pearson's $r = - 0.9$ with $p=0.025_{-0.012}^{+0.23}$. The data are inconsistent with no correlation between disk mass and FUV irradiation. 

We can now empirically fit the relation between median disk mass and FUV radiation field strength for the first time. We find $M_{\rm{dust,median}} = -1.3^{+0.14}_{-0.13} \log_{10}(F_{\rm{FUV}} / G_0) + 5.2^{+0.18}_{-0.19}$. Surprisingly, even radiation fields of $\sim 10\,G_0$ appear to affect disk mass, at the age of the Orion A disks ($\sim 1 - 3$\,Myr). In the most irradiated environments probed here, disk masses have dropped by a factor 2 after this time. Once again, the most massive disks show a similar trend, with the 95th percentile of most massive disks dropping below $25\,M_{\oplus}$ where $F_{\rm{FUV}} > 40 G_0$.

Figure~\ref{fig:M_vs_G0} also shows the data for $\sigma$ Ori~\citep{ansdell17} and the ONC~\citep{eisner18}. Strikingly, $\sigma$ Ori is not inconsistent with the trend. The ONC, on the other hand, has significantly higher disk masses, restating the proplyd lifetime problem, but see~\citet{winter19}.

\section{Discussion}
Having found a loss of disk mass with proximity to B-type and A0 stars in the southern part of Orion A opens the door to studying the link between FUV irradiation and disk evolution more generally. Two key questions are the universality and the physics behind the observed trend. We discuss these questions here.

\subsection{Disk mass and FUV irradiation across the survey area}
\label{sec:zoom}
\begin{figure*}[h]
    \centering
    \includegraphics[width=\linewidth]{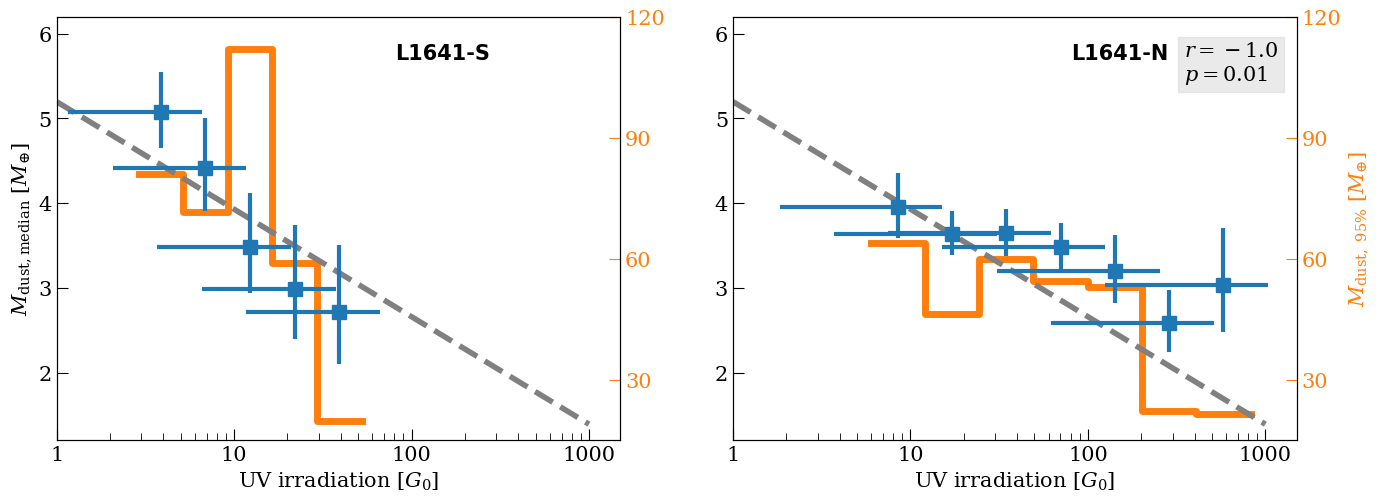}
    \caption{Disk masses binned by incident FUV flux (in units of $G_0$) for L1641-S (defined as disks between $211.5^{\circ} \ell 213.5^{\circ}$) and L1641-N ($\ell > 211.5^{\circ}$). Blue points show median dust masses in each bin, with uncertainties; the empirical 95th percentile disk mass is shown in orange. The dashed lines show the empirical relationship derived for the full sample in Fig.~\ref{fig:M_vs_G0}.}
    \label{fig:M_vs_G0_ZOOM}
\end{figure*}

As shown in Figure~\ref{fig:G0map}, L1641-N and L1641-S contain the most irradiated disks in the sample. In Figure~\ref{fig:M_vs_G0_ZOOM}, we separate these two subsamples. L1641-S contains stars between $211.5^{\circ} < \ell \leq 213.5^{\circ}$, while L1641-N has all disks in the SODA catalog with $\ell > 211.5^{\circ}$. Both regions are consistent with the empirical trend derived for the full sample. The limited range of $F_{\rm{FUV}}$ sampled in L1641-S makes it difficult to statistically identify a trend in median disk mass. The increase in 95th percentile disk mass at $10\,G_0$ in L1641-S is notable. This subsample is also responsible for an increase in 95th percentile disk mass in Fig. 2 around 0.6-1\,pc. This is not seen in the median disk mass, so the responsible sources are outliers, related to younger populations to either side of L1641-S (see~\citet{SODA}, Fig. 9).

In L1641-N, the trend is significant. In the more irradiated regimes in L1641-N, the disk mass (both median and 95th percentile) may be higher than elsewhere in L1641. This could be a consequence of the possibly complicated age-distance structure in L1641-N: ~\citet{SODA} found a lower disk mass caused by an older (foreground) population in the direction of L1641-N as a whole, potentially linked to the location of the older NGC 1980 cluster~\citep{alves12,zari19}. However, this older population extends over a much larger region than the 0.5\,pc sphere-of-influence of most A0 and B-type stars in the sample.

\subsection{Driving local disk mass loss in L1641 and L1647}
After correcting for larger-scale trends in stellar age (and, thus, disk mass), we observe that FUV field strength is correlated with decreasing disk mass. While age is clearly the primary driver of disk mass evolution, FUV irradiation is an important secondary effect. This conclusion is strengthened because our result does not rely on data for a single ionizing star, as discussed in the previous paragraph. To the extent that L1641-N is a less pristine sample, we should expect older disks to be less massive overall, and weaken the imprint of external photoevaporation. This may explain the shape of the trend beyond $F_{FUV} > 100\,G_0$, although this is difficult to say without a larger sample in this irradiation regime.

Apart from large-scale age gradients, a second confounding effect is the external heating of disks near massive stars. This was simulated for a representative sample of disks in~\citet{haworth21}. Even in the radiation regimes discussed here, more irradiated disks are hotter than the 20\,K effective dust temperature we assumed. This is further enhanced by the truncation of the most irradiated disks due to the inside-out disk destruction in FUV-dominated regimes~\citep[e.g.][]{FRIED, eisner18}. The net consequence of this is again a flattening of the relation between disk mass and FUV radiation field.

Likewise, we can only plot median disk masses for stars which still bear disks. If the disk is lost completely -- either due to internal evolution alone, or in combination with external irradiation-driven mass loss -- we did not include the star in our targets. This lowers our sensitivity, although this is compensated to some extent by the fact that inner disks are difficult to destroy in these radiation regimes~\citep{FRIED}.

So far, we ignored interstellar absorption and the three-dimensional structure of the cloud. While the Class II disks in this sample are longer significantly shielded by envelopes, they are close to the cloud (see, for instance, the discussion in~\citet{kainulainen17}). This, as well as the large uncertainties in the 3D separation between the disks and the A0 and B-type stars, will only increase the scatter in each bin, and cannot drive the trend we observe.

What constraints does ALMA place on the physics behind disk destruction? We observed the continuum emission from millimeter-sized dust grains in these disks, which are seen to be located in the disk midplane at very low scale heights~\citep[e.g.,][]{villenave20, villenave22}. External irradiation will preferentially act on the surface layers and outer parts of the disk, which are not well-traced by these grains. However, millimeter continuum emission is tightly correlated with the radial extent of the continuum emission~\citep[e.g.][]{tripathi17,andrews18}. If external photoevaporation primarily drives an outside-in disk destruction, we can understand our observations as being caused a shrinking disk radius at millimeter wavelengths. Ultimately, resolved ALMA observations of the dust and the bulk gas, as traced by $^{12}$CO and CI in these irradiated environments are needed to definitively understand the interaction between UV-driven mass loss and dust particles in these disks.

The disk population in the $\sigma$ Ori SFR is probably the most comparable to the disks studied here. Its radiation field is primarily due to a single O9 star ($\sigma$ Ori Aa), and as such more uniform and somewhat stronger than the environments studied here, while the disk population is likely to be slightly older ($3-5$\,Myr; \citealp{oliveira04}) than ours ($1-3$\,Myr; \citealp{dario16}). The sample is too small to follow the evolution of the median disk mass, but Figure 6 of~\citet{ansdell17} suggests the same behavior reported here, with the detection fraction and mass of the most massive disks dropping for separations $<2$\,pc. As shown in Figure~\ref{fig:M_vs_G0}, averaging over the region, $\sigma$ Ori supports the conclusion that FUV-driven disk mass loss has a measurable impact on disks in the vicinity even of A0 and B-type stars.

\section{Conclusions}
The impact of external FUV irradiation on the evolution of protoplanetary disks provides important constraints on the potential for planet formation in many stars in the galaxy. Using the SODA survey~\citep{SODA}, which contains a large (N=873) sample of Class II disks in Orion, in a variety of intermediate irradiation regimes ($1 - 1000\,G_0$), we can for the first time study the evolution of the median disk dust mass, as traced by ALMA continuum observations.

Our data support the interpretation that FUV irradiation can -- even in this regime -- significantly affect the evolution of disks, with disks losing a factor $\sim 2$ in mass over two orders of magnitude in FUV field strength, and we can for the first time fit an empirical relation between disk mass and irradiation. While the environments studied here are much less extreme than e.g. the ONC, A0 and B-type stars for the are much more common, and similar radiation fields are encountered by a large fraction of young stars even in the Solar neighborhood~\citep{winter22}. This emphasizes the need for planet population syntheses to take this process into account, as well as resolved studies of the disks in these radiation environments, in order to link the observed exoplanet population to the well-studied nearby protoplanetary disks.

\begin{acknowledgements}
    We thank the anonymous referee for insightful comments which helped improve the quality of this Letter. This paper makes use of the following ALMA data:
ADS/JAO.ALMA\#2019.1.01813.S. ALMA is a partnership of ESO (representing its member states), NSF (USA) and NINS (Japan), together with NRC (Canada), MOST and ASIAA (Taiwan), and KASI (Republic of Korea), in cooperation with the Republic of Chile. The Joint ALMA Observatory is operated by ESO, AUI/NRAO and NAOJ. This project has received funding from the European Research Council (ERC) under the European Union’s Horizon 2020 research and innovation programme (Grant agreement No. 851435)

\end{acknowledgements}

\bibliographystyle{aa}
\bibliography{sodaii_v1}

%
%

\end{document}